\documentclass[aps,pra,preprint,amsmath,superscriptaddress]{revtex4}

\usepackage{graphicx}

\begin{document}

% {
% \large\bf\noindent
% \underline{Report to Science:}\\[1em]
% }
\begin{center}
{
\large\bf\noindent
A new intermolecular mechanism to selectively drive\\ photoinduced damages\\[1em]
}
{
\bf\noindent
Kirill Gokhberg$^1$, P\v{r}emysl Koloren\v{c}$^2$, Alexander I. Kuleff$^1$, Lorenz S. Cederbaum$^1$\\
}
{
\em\noindent
$^1$Theoretische Chemie, Physikalisch-Chemisches Institut, \\ 
Universit\"{a}t Heidelberg, Im Neuenheimer Feld 229, D-69120 Heidelberg, Germany\\[0.5em]
\noindent
$^2$Institute of Theoretical Physics, Faculty of Mathematics and Physics, \\
Charles University in Prague, V Hole\v{s}ovi\v{c}k\'{a}ch 2, 180 00, Prague, Czech Republic\\[1.5em] 
}
\end{center}

\textbf{Abstract:}
Low-energy electrons (LEEs) are known to be effective in causing strand breaks in DNA. Recent experiments show that an important direct source of LEEs is the intermolecular Coulombic decay (ICD) process. Here we propose a new cascade mechanism initiated by core excitation and terminated by ICD and demonstrate its properties. Explicit calculations show that the energies of the emitted ICD-electrons can be controlled by selecting the initial atomic excitation. The properties of the cascade may have interesting applications in the fields of electron spectroscopy and radiation damage. Initiating such a cascade by resonant X-ray absorption from a high-Z element embedded in a cancerous cell nucleus, ICD will deliver genotoxic particles \textit{locally} at the absorption site, increasing in that way the controllability of the induced damage.
\\[1em]

When embedded in a medium, electronically excited atoms and molecules efficiently decay radiationlessly by transferring their excess energy to the neighboring species in the environment and ionizing them, creating in that way low-energy electrons (LEEs) and radical cations. This process is known as intermolecular Coulombic decay (ICD) \cite{cederbaum1997}.

Since its discovery in 1997 \cite{cederbaum1997}, the ICD has been successfully investigated in a variety of systems \cite{Uwe_review}. It usually proceeds on a femtosecond timescale and becomes faster the more neighbors are present, dominating most of the competing relaxation processes. Experimental investigation of ICD in water dimers \cite{jahnke2010waterdimer} found the rate of this process to be so large as to completely suppress the proton transfer in the inner-valence ionized water molecules. As a result of ICD, two intact water cations are produced by the consecutive Coulomb explosion in addition to the slow ICD electrons. Another striking feature of this process is that ICD remains effective for considerable interatomic distances. ICD was demonstrated experimentally and theoretically for He dimer, which is the weakest bound system known in nature, and found to be operative over distances of about 45 times the  atomic radius \cite{Sisourat10_1,Havermeier10_1}.

These properties of ICD and the fact that it appeared to be ubiquitious in hydrogen-bonded systems \cite{Mueller06,Schwartz10,Stoychev11} suggest its potential importance for radiation damage \cite{Howell_rev}. Low-energy electrons (LEEs) \cite{Boudaiffa03032000,doi:10.1021/ja029527x}, as well as radical cations \cite{vonSonntag}, the direct products of the ICD, are known to be effective in causing single- and double-strand breaks in DNA. Recent experiments even suggest that ICD electrons contribute up to about 50\% of the single-strand breaks (SSB) in DNA \cite{Grieves11}. Moreover, these electrons possess energies at which much harder to repair double-strand breaks (DSB) occur \cite{Boudaiffa03032000,Mucke10} and the energetic cations produced in Coulomb explosions may additionally damage the DNA. 

The excited electronic states undergoing ICD with the environment may be produced directly by photoabsorption, electron impact, or even by ion impact as demonstrated recently \cite{Kim11}. Alternatively, they may be formed as a result of multistage cascade processes. The Auger decay process followed by ICD is one type of such cascades. It is initiated by core ionization of an atom, e.g., through X-ray absorption. This cascade, postulated theoretically \cite{santra2003}, has since been studied in a series of experimental works \cite{morishita2006,yao_PRL2008,Kreidi09}. Importantly, since the Auger-ICD cascade is initiated by core ionization, in a complex system one has little control over the location where the Auger decay can be initiated and the follow-up ICD is going to take place. Indeed, in a polyatomic system, all atoms with core-ionization potentials below the energy of the impacting photon may become ionized and, therefore, undergo an Auger transition. Here we propose a different scheme to initiate a cascade ending by ICD in which one has control not only over the location of the process, but also over the energies of the emitted ICD electrons.

If the energy of the incoming photon lies just below the core-ionization threshold of a selected atom in a larger system, at a number of discrete energies the core electron will resonantly absorb the photon and be promoted to some bound unoccupied orbital. The resulting highly energetic core-excited state may decay through the emission of an Auger electron in the process known as resonant Auger (RA) decay \cite{Aksela1999,Piancastelli,Hergenhahn2003}. Here, a valence electron fills the initial vacancy and another valence electron is ejected into the continuum, while the initially excited electron remains a spectator. This commonly termed spectator RA mechanism produces highly excited valence-ionized states (so-called photoionization satellite states). The alternative participator process, in which the initially excited electron participates in the decay, is usually the much less efficient de-excitation pathway following core excitations \cite{FP1990115}.

Using modern high-resolution synchrotron-radiation sources one can selectively excite core electrons not only on chemically different atoms but also on identical atoms occupying non-equivalent sites in the system. The latter stems from the different chemical shifts the atoms experience in a different chemical environment. This selectivity is used in the Near Edge X-ray Absorption Fine Structure spectroscopy (NEXAFS) \cite{bk:nexafs} to study, for example, the bonding in biologically relevant organic molecules \cite{doi:10.1021/jp803017y}.

\begin{figure}[ht]
\begin{center}
\includegraphics[width=8.5cm]{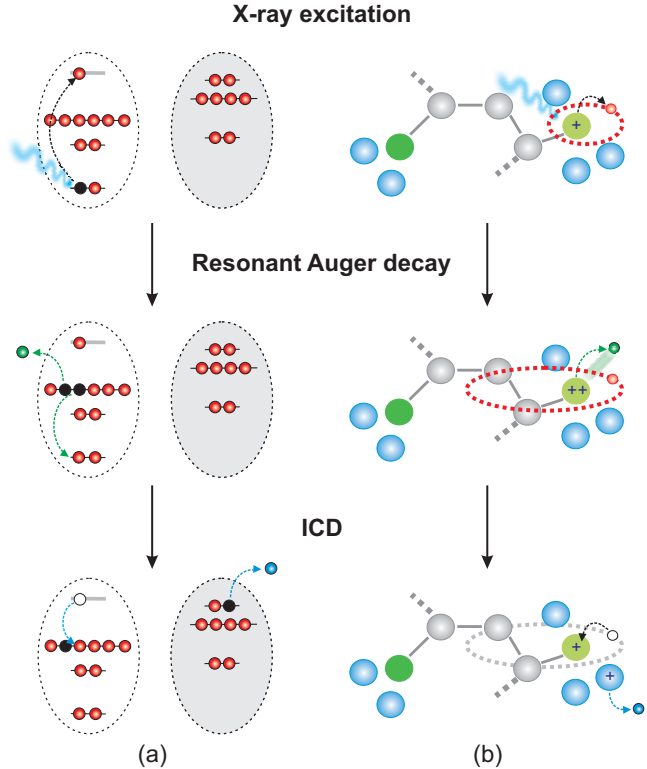} 
\caption{Schematic illustration of the resonant Auger-ICD cascade. (Panel a) The mechanism. A parent core excited state embedded in the environment decays locally in the spectator resonant Auger process producing ionized-excited states. The latter continue decaying in the ICD process ionizing the neighbors in the environment. The two cations produced by ICD repel each other strongly and undergo a Coulomb explosion (not shown). (Panel b) Selectivity property. The parent state is produced selectively on a given atom of the embedded system. The ionized-excited states formed in the RA process tend to be localized close to the site of the initial excitation and decay by ICD ionizing predominantly neighbors from the environment nearest to this site.}
\label{fg2}
\end{center}
\end{figure}

The RA decay takes place on the initially excited species, and the excited ionic states produced usually have excess energies of a few tens of eV which they can transfer efficiently to the environment by continuing to decay electronically via ICD. We illustrate schematically the resulting RA-ICD electronic cascade in Fig.~\ref{fg2}. It bears decisive differences from the Auger-ICD cascade described above, allowing one to control the ICD process. First, the energy of emitted ICD electrons in the \textit{same} environment depends sensitively on the energies and population of the states produced by RA. These parameters depend in turn on the nature of the parent core-excited state. Consequently, by varying the energy of the high-energy photon one can resonantly excite different parent states and change the appearance of ICD spectra in a controlled manner. Second, the initial parent core excitation can be placed selectively on an atom in a moiety of choice. And as the RA decay tends to proceed locally populating ionized-excited states with two holes localized predominantly on the atom bearing the initial excitation, see e.g. Ref.~\cite{Hergenhahn2003}, ICD will follow leading mostly to the ionization of the environment in the vicinity of the parent core-excitation (see Fig.~\ref{fg2}b). In other words, the site where the damaging ICD electrons are produced can be selectively chosen.

We illustrate the RA-ICD cascade on the example of ArKr. The relative simplicity of this system gives a transparent picture of the processes involved. This cascade can be initiated by selectively producing a core-excited state localized on the Ar atom. Choosing a photon energy of 246.51~eV one populates the $2p^{-1}_{1/2}4s$ state of Ar \cite{Gouw1995}. This state lives only 5.5~fs \cite{PhysRevA.22.1615} and decays locally by spectator Auger populating a band of excited states of Ar$^{+}$ (see the Supplementary Materials). These states lie at energies between 17 and 22~eV above the ground state of Ar$^{+}$ and can, therefore, undergo ICD with the neighboring Kr whose lowest ionization potential is 14~eV. To demonstrate that these states indeed further decay by ICD, we have determined the ICD rates employing extensive \textit{ab initio} many-body calculations. 

\begin{figure}[ht]
\begin{center}
\includegraphics[width=8.5cm]{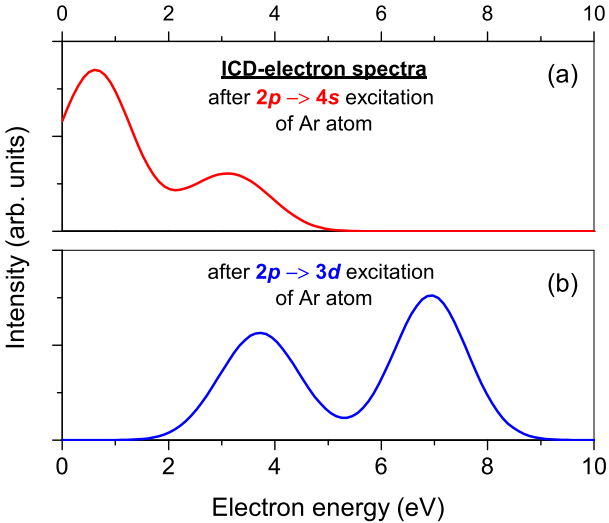}
\caption{Spectra of the ICD electrons emitted in the RA-ICD cascades in ArKr: (a) following core-excitation of the Ar$(2p^{-1}_{1/2}4s)$ parent state at 246.51~eV, and (b) following the core-excitation of the Ar$(2p^{-1}_{3/2}3d)$ parent state at 246.93~eV. One sees that two different core-excitations of the same atom lead to totally different energy distributions of the ICD electrons. The possibility to control the energies of the ICD electrons becomes apparent. For more details, see text and the Supplementary Materials.} 
\label{fg4}
\end{center}
\end{figure}

We can estimate the spectra of the ICD electrons emitted in the cascade using the computational scheme presented in the Supplementary Materials. The corresponding electron spectrum is depicted in Fig.~\ref{fg4}a. It exhibits two peaks: a  pronounced peak between 0 and 1~eV, and a weaker peak between 2 and 4~eV. Following the ICD, Ar$^{+}$ and Kr$^{+}$ will repel each other resulting in the Coulomb explosion. At the end of this dissociative process, the ions acquire $\sim3.7$~eV kinetic energy.

Increasing the energy of the X-ray photon by just 0.4~eV to 246.93 eV one excites the $2p^{-1}_{3/2}3d$ parent state of Ar. The RA decay of this core excitation populates a totally different band of excited states of Ar$^+$. All these states can further decay via ICD emitting electrons whose spectrum is shown in Fig.~\ref{fg4}b. It consists again of two peaks: one between 3 and 5~eV and another between 6 and 8~eV. One sees that two different core-excitations of the same atom lead to totally different energy distributions of the ICD electrons. The possibility to control the energies of the ICD electrons becomes apparent.

While details may differ, the mechanism of RA-ICD cascade in other systems will be similar to the case of ArKr. Very recently, the RA-ICD cascade was demonstrated experimentally in molecular dimers \cite{Till}. The selectivity property of the cascade and the ability to control the energies of ICD electrons by tuning to the different parent state make RA-ICD cascade a foundation for a promising analytical technique. For example, in a larger molecule embedded in a solvent one may create a core excitation localized on a selected moiety. Auger decay is an intra-molecular process and by observing the Auger electrons one studies the electronic structure of the molecule at the excitation site. ICD is an intermolecular process and involves the neighbors, and the observation of ICD electrons allows one to probe the local environment (see Fig.~\ref{fg2}b). 

\begin{figure}[htb]
\begin{center}
\includegraphics[width=8.5cm]{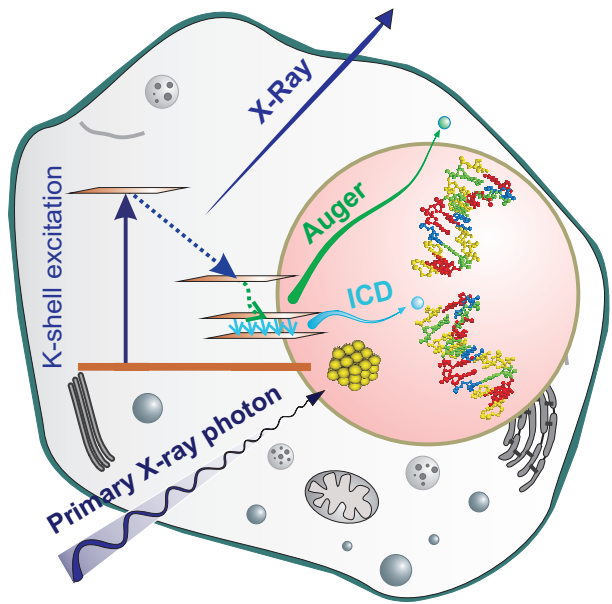}
\caption{Schematic illustration of the proposed biomedical application of the RA-ICD cascade. A monochromatic high-energy X-ray photon is resonantly absorbed by a K-shell electron to produce a core-excited parent state of a high-Z compound brought to the vicinity of the DNA of the cell to be damaged. Due to the high energy, these X-rays have a large penetration depth in the tissue. The use of high-Z compounds in combination with the resonant core-excitation also has the advantages of improving the contrast between the healthy and cancerous tissue enabling to use lower concentrations of such --- often toxic --- compounds, and lower intensity of the X-rays compared to the case where less specific core-ionization is employed. The parent state decays in a local multistage cascade of X-ray emission and resonant Auger processes emitting secondary photons and energetic Auger electrons, and producing thereby ionized-excited, doubly ionized-excited, etc. states of the moiety containing the high-Z element. These states continue to decay intermolecularly by ICD releasing genotoxic low-energy electrons and causing additional mechanical damage to DNA by Coulomb explosions.} 
\label{fg5}
\end{center}
\end{figure}

Let us elaborate on another possible practical application of the high selectivity and controllability of RA-ICD and related cascades (for brevity we will call them all RA-ICD). The properties of these cascades and the fact that they can be initiated by high-energy X-rays makes them potentially useful in controlling the radiation damage of living cells. Embedding a high-Z element in the nucleus of a cancerous cell, a high-energy photon tuned resonantly to a particular core-excited state of this element will be predominantly absorbed by this element and trigger a RA-ICD cascade (see Fig.~\ref{fg5}). In contrast to the cascade in the low-Z ArKr system, where one Auger and one ICD electron are emitted, a K-shell excitation of a high-Z element will usually initiate a complex multistage process. Its first steps will be dominated by fluorescence \cite{RevModPhys.44.716}. The later steps will proceed radiationlessly by emitting either Auger electrons or ICD electrons in the so-called core-ICD process \cite{Aziz_Nature08,Pokapanich_JACS11} (see also the Supplementary Materials). The final states of the decay will still be ionized-excited states which will continue decaying by ICD with the water shell surrounding the DNA or with the DNA itself. 

We note that radiotherapeutic techniques based on DNA-incorporated high-Z Auger-electron emitters (radionuclides \cite{Adelstein_CBR2003} or photon activated ones \cite{fairchild1982}) have already been suggested. For instance, platinum-containing or iodinated compounds such as iodo-deoxyuridine has been extensively investigated. The damage is thought to arise primarily from two sources. The Auger cascade initiated in the high-Z element either spontaneously (in radionuclides) or by absorbing an X-ray photon leads to the emission of genotoxic Auger electrons with energies below 500~eV \cite{Pomplun_IJRB91}. Following this cascade, electron transfer to the highly charged high-Z ion from the environment may also result in a Coulomb explosion causing further damage \cite{KuemPomplun_JMS10}. In the proposed RA-ICD cascade, in addition to the Auger and the above mentioned core-ICD electrons, highly damaging low-energy ICD electrons with \textit{controllable} energies are  emitted. For example, SSBs in DNA are produced favorably by electrons with energies between 0 and 4~eV \cite{PhysRevLett.93.068101}, while DSBs are mostly induced by electrons with energies above 6~eV \cite{doi:10.1021/ja029527x}. We note that only for electrons with energies below 15~eV the microscopic mechanisms for strand breakage have been investigated \cite{Boudaiffa03032000,doi:10.1021/ja029527x,Simons_ACR06,Sanche_PRE07,Orlando_IJMS08} (for brief discussion on higher energy electrons, see Supplementary Materials). In all the ICD processes during the cascade two or more neighboring ions are produced \textit{directly} leading to damaging Coulomb explosion \cite{jahnke2010waterdimer,Oriol_CPC10}. The damage to DNA through the ICD electrons and the Coulomb explosion following RA-ICD cascades will happen in the \textit{immediate vicinity} of the site where the energy was initially deposited.

Additional benefits of the proposed scheme is that the RA-ICD cascade is triggered by resonant photon absorption which is more efficient than the traditional photon activated techniques where the Auger cascade is initiated by K-shell ionization.  As we saw in the ArKr example, the RA-ICD cascade provides the opportunity to tune the energies of the slow electrons by using the site and energy selectivity of the resonant core-excitations process which, in turn, may be useful to increase the damage of unwanted cells. Very importantly, in each ICD process, including the core-ICD ones which may take place at each step of the cascade, a genotoxic electron and a radical cation are \textit{simultaneously} produced. The latter are also known to be extremely effective in causing DNA lesions \cite{vonSonntag}.

Understanding the microscopic mechanisms of DNA lesions would enable us to find the relevant parameters needed to increase the control over the induced damage and to maximize it, paving the way for efficient cancer therapies.

% \vspace{2em}
% 
% \noindent{\bf Methods}
% 
% The ICD lifetimes of the involved states were computed using an \textit{ab initio} many-body method \cite{averbukh2005}. The method is based on the general Fano resonance formalism in which the initial decaying state is represented as a bound (discrete) state embedded in the continuum of final states of the decay. The ${\cal L}^2$ approximations for the discrete and continuum components of the $(N-1)$-electron wave function are obtained within the Green's function ADC approach \cite{nd-ADC98} and the resulting discretized spectrum is renormalized and interpolated in energy using the Stieltjes imaging technique. The potential energy curves of the initial and final ICD states were modeled using atomic data. The final ICD-electron spectra were obtained by convoluting the discrete electronic transitions with an appropriate Gaussian profile. For more details, see the Supplementary information.

\bibliographystyle{Science}
%\bibliography{lit_full} 

\vspace{2em}

\noindent{\bf Acknowledgments:} The research leading to these results has received funding from the European Research Council under the European Community's Seventh Framework Programme (FP7/2007-2013) / ERC Advanced Investigator Grant n$^\circ$ 227597.

\vspace{2em}

\noindent{\bf Supplementary Materials}

\noindent Materials and Methods\\ Supplemantary Text\\ Figures S1 to S3\\ References (XX--XX)

\vspace{2em}

\noindent{\bf Authors contribution}

\noindent K.G., A.I.K., and L.S.C. conceived the cascade mechanism and its potential consequences. P.K. computed the lifetimes, and K.G. and A.I.K. evaluated the electronic spectra. K.G., A.I.K., and L.S.C. wrote the paper.

% \vspace{2em}

\end{document}